\documentclass[12pt]{article}
\usepackage{mathptmx}      
\usepackage{graphicx}
\usepackage{mathtools}
\usepackage{wrapfig}
\usepackage{caption}
\usepackage{authblk}
\usepackage{geometry}
\newcommand{\bra}[1]{\langle #1|}
\newcommand{\ket}[1]{|#1\rangle}

\newcommand{\half}{\frac{1}{2}}

\newcommand{\ex}[1]{\mathrm{e}^{#1}} 
\newcommand{\abs}[1]{\left|#1\right|}

\title{Experimental state control by fast non-Abelian holonomic gates with a superconducting qutrit}
\author[]{S. Danilin}
\author[]{A. Veps\"al\"ainen}
\author[]{G.~S. Paraoanu }
\affil[]{Low Temperature Laboratory, Department of Applied Physics, Aalto University School of Science, P.O. Box 15100, FI-00076 AALTO, Finland}

\begin{document}
\begin{titlepage}
\maketitle






\begin{abstract}
	Quantum state manipulation with gates based on geometric phases acquired during cyclic operations promises inherent fault-tolerance and resilience to local fluctuations in the control parameters. Here we create a general non-Abelian and non-adiabatic holonomic gate acting in the $(\ket{0},\ket{2})$ subspace of a three-level transmon fabricated in a fully coplanar design. Experimentally, this is realized by simultaneously coupling the first two transitions by microwave pulses with amplitudes and phases defined such that the condition of parallel transport is fulfilled. We demonstrate the creation of arbitrary superpositions in this subspace by changing the amplitudes of the pulses and the relative phase between them. We use two-photon pulses acting in the holonomic subspace to reveal the coherence of the state created by the geometric gate pulses and to prepare different superposition states. 
	We also test the action of holonomic NOT and Hadamard gates on superpositions in the $(\ket{0},\ket{2})$ subspace. 

\end{abstract}

\thispagestyle{empty}
\end{titlepage}
\newpage

\section{Introduction}
\label{intro}
The creation of robust and fast quantum gates is one of the most important problems in the field of quantum computation and quantum information processing. The presence of noise and relaxation makes this problem quite difficult in practice. One possible solution is to use geometric phases \cite{Pancharatnam,Berry,Wilczek,Aharonov,Anandan}: under cyclic evolution, the geometric gates depend only on global features of the path the system undergoes in the space of control parameters, which endows them with intrinsic noise-tolerance against local fluctuations in these parameters. Moreover, it was shown that the gates based on nonabelian geometric phases can form a universal set, thus enabling quantum computation \cite{Zanardi}. This approach is known as holonomic (or geometric) quantum computation. 

One type of scheme for holonomic quantum computation relies on adiabatic operations: this was proposed for systems of trapped ions \cite{Duan}, superconducting nanocircuits based on Josephson junctions \cite{Faoro,Pirkkalainen}, semiconductor quantum dots \cite{Solinas}, superconducting qubits \cite{Kamleitner}, and demonstrated experimentally using a nuclear magnetic resonance \cite{Jones}, and electron spin resonance \cite{Wu}. The disadvantage of adiabatic schemes is their long duration, which still exposes the system to noise. Several strategies for shortening this time are currently pursued; for example, with superconducting qubits the superadiabatic acceleration of the stimulated Raman adiabatic passage \cite{stirap} has been recently demonstrated \cite{sastirap}. In the case of Abelian phases, it has been known for some time \cite{Aharonov} how to construct non-adiabatic geometric phases \cite{Wang}. However, universal holonomic quantum computation cannot be realized with the use of only Abelian (commuting) operations. To construct fast non-Abelian phases, instead of using slow control in the full Hilbert space, the idea is to use only a subspace of the Hilbert space under a cyclic evolution of the control parameters \cite{Sjoqvist}.  
Following this proposal, non-adiabatic and non-Abelian holomonic gates were demonstrated experimentally with a liquid NMR quantum information processor \cite{Feng}, a superconducting three-level atom in a 3D cavity \cite{Abdumalikov}, and in solid-state spins of diamond nitrogen-vacancy centres \cite{Arrojo-Camejo,Zu,Yale,Sekiguchi}.
 
Differently from the previous experiments, where the focus was the characterization of a discrete set of standard qubit gates obtained by holonomic operations, here we demonstrate the preparation of arbitrary quantum superpositions under general single-qubit fast non-Abelian holonomic gates (holonomic rotations). Also, in view of the goal of realizing a multi-qubit holonomic processor, we work in a 2D-coplanar architecture, which allows straightforward scalability. Although the decoherence in these architectures is higher than in the 3D case, our results show that the holonomic operations can be realized. The device used is a transmon \cite{Koch}, and the holonomic gates are implemented by using two microwave gates with externally-controlled amplitudes and phases. To demonstrate that the states prepared are coherent, we use a two-photon pulse which directly couples the states $\ket{0}$ and $\ket{2}$. Finally, by changing the order of the two-photon and holonomic pulses, we investigate the action of holonomic NOT and Hadamard gates on superpositions of states in the $(\ket{0},\ket{2})$ subspace. The advantage of using a two-photon pulse rather than a sequence of pulses $0 - 1$ and $1 - 2$ is that it shortens the overall operation time, which is favorable in reducing the effects of decoherence.
    
\section{Holonomic gates}

In general the phase that the quantum system accumulates during its evolution consists of two components: a dynamical part, which is given by the time-integral of the instantaneous energy eigenvalue, and a geometric part, which depends on the path followed by the system in a space defined by the control parameters. If the evolution is cyclic, the geometric phase accumulated depends only on global features of the path (the encompassed solid angle), which potentially renders this phase insensitive to certain local  noises \cite{Wu,Leek,Filipp}. 

\begin{figure}[h!]
	\centering
	\includegraphics[width=\textwidth]{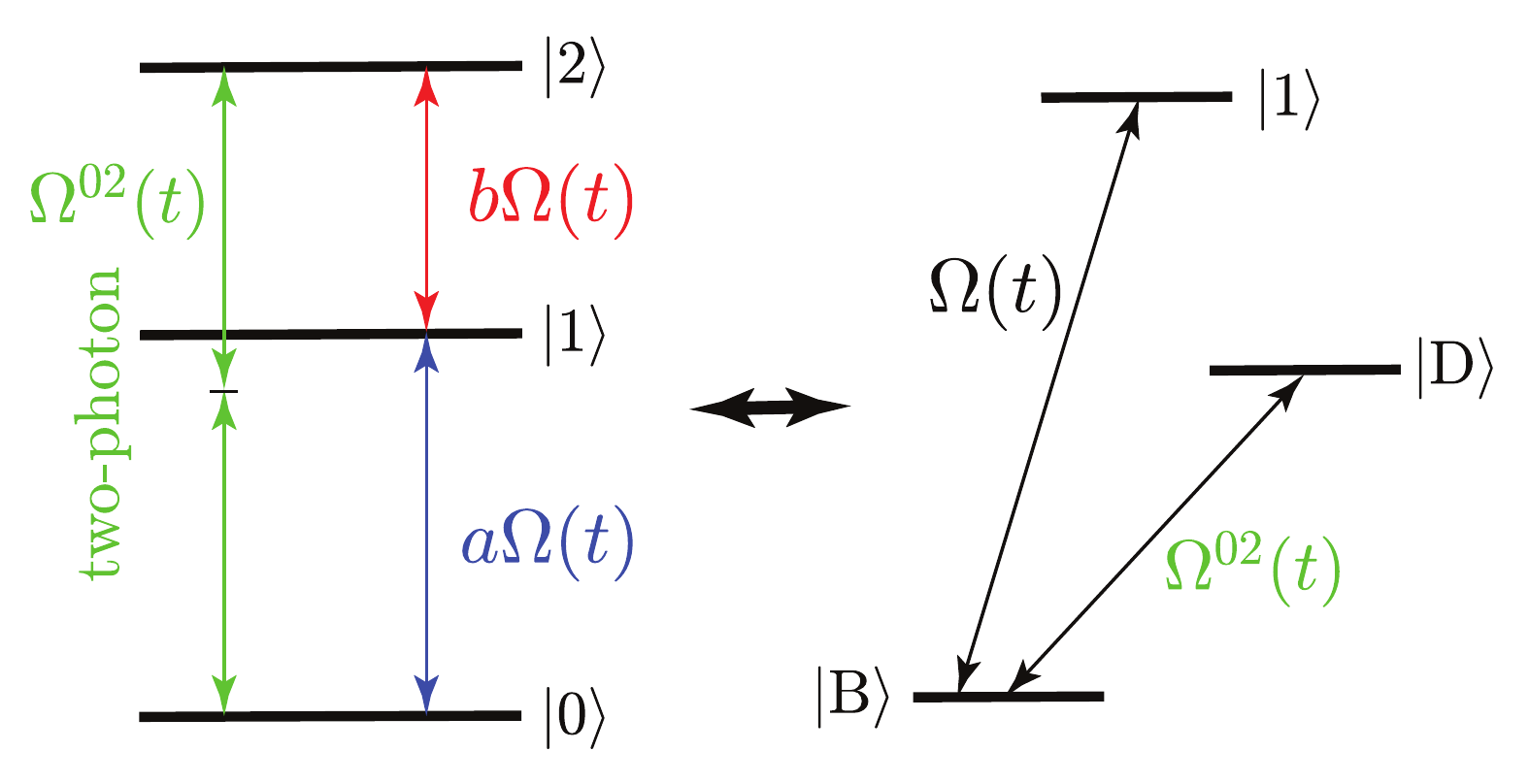}
	\caption{Actual three-level system (left diagram) with the two drives used to create the holonomic gates applied to the $0 - 1$ and $1 - 2$ transitions. The $0 - 2$ transition is driven via a two-photon process. The change of basis from $(\ket{0},\ket{1},\ket{2})$ to $(\ket{\textrm{D}},\ket{1},\ket{\textrm{B}})$ can be visualized  by the level diagram presented on the right. The dark state decouples completely during the holonomic operations and it is coupled only by the two-photon drive (right diagram).}
	\label{three_levels}
\end{figure}

Following the proposal of Sj\"{o}qvist et al. \cite{Sjoqvist} and its experimental realizations in several systems \cite{Feng,Abdumalikov,Arrojo-Camejo,Zu}, we implement fast (non-adiabatic) and non-Abelian geometric gates by applying simultaneous drives $a\Omega(t)$ and $b\Omega(t)$ to the $0 - 1$ and $1 - 2$ transitions of a three-level superconducting artificial atom (Fig. \ref{three_levels} left panel). The pulses have identical shapes $\Omega(t)$, but different amplitudes controlled by the scaling parameters $\abs{a}$ and $\abs{b}$. The states $\ket{0}$ and $\ket{2}$ form a computational basis, while the state $\ket{1}$ acts as an auxiliary state, and ends up unpopulated at the end of the gate.

The Hamiltonian of the drive in a double-rotating frame and within the rotating-wave approximation (see eg. \cite{stirap,photonics,Jian_prb}) reads
\begin{equation}
\label{hamiltonian_012}
\hat{H}(t)=\frac{\hbar\Omega(t)}{2}(a\ket{1}\bra{0}+b\ket{1}\bra{2}+h.c.),
\end{equation}  
where for convenience we take $\abs{a}^2+\abs{b}^2=1$. We 
introduce the "dark" $\ket{\textrm{D}}=-b\ket{0}+a\ket{2}$ and the "bright" $\ket{\textrm{B}}=a^*\ket{0}+b^*\ket{2}$ states, which define an orthonormal basis in the subspace $(\ket{0},\ket{2})$. In terms of these states, the Hamiltonian~(\ref{hamiltonian_012}) can be written as 
\begin{equation}
\hat{H}(t)=\frac{\hbar\Omega(t)}{2}(\ket{1}\bra{B}+\ket{B}\bra{1}) = \frac{\hbar\Omega(t)}{2}\sigma_x^{(1,\textrm{B})},
\label{hamiltonian_D1B}
\end{equation}
where $\sigma_x^{(1,\textrm{B})}$ defines a Pauli-$x$ operator 
in the subspace $\{|1\rangle , |B\rangle \}$. Note that 
the "dark" state $\ket{\textrm{D}}$ is decoupled from the drives (see Fig. \ref{three_levels}).

The basis vectors $(\ket{0},\ket{1},\ket{2})$ evolve then as  $\ket{\psi_j(t)}= \exp\left(-\frac{i}{\hbar}\int_0^t{\hat{H}(\tau)d\tau}\right)\ket{j},\\ (j=0,1,2)$. In the case of non-adiabatic evolution, these are not instantaneous eigenstates of $\hat{H}(t)$. Nevertheless the parallel transport condition (absence of the transitions) $\bra{\psi_i(t)}\hat{H}(t)\ket{\psi_j(t)}=0$ for the states $((i,j)=(0,2),i\ne j)$ is fulfilled if the parameters $a$ and $b$ are kept constant during the gate operation, and thus the gate realized in this way is purely geometric \cite{Abdumalikov}. 

If the drives are chosen so that the evolution is cyclic in the $(\ket{1},\ket{\textrm{B}})$ subspace, $\int_0^t{\Omega(\tau)d\tau}= 2\pi$, the time evolution unitary operator in the basis $(\ket{\textrm{D}},\ket{1},\ket{\textrm{B}})$ can be expressed as
\begin{gather}
\hat{\tilde{U}} = \ket{\textrm{D}}\bra{\textrm{D}}+\exp{\left(-\frac{i}{2}\int_0^t{\Omega(\tau)d\tau\sigma_x^{(1,\textrm{B})}}\right)}\\
=\ket{\textrm{D}}\bra{\textrm{D}} + \hat{I}^{(1,\textrm{B})}\cos{\pi}-i\hat{\sigma}_x^{(1,\textrm{B})}\sin{\pi}\\ =\ket{\textrm{D}}\bra{\textrm{D}}-\ket{1}\bra{1}-\ket{\textrm{B}}\bra{\textrm{B}}\\
=\begin{pmatrix}
1&0&0\\0&-1&0\\0&0&-1
\end{pmatrix},
\label{unitary_D1B}
\end{gather}
where $\hat{I}^{(1,\textrm{B})}=\ket{1}\bra{1}+\ket{\textrm{B}}\bra{\textrm{B}}$.
The first term $\ket{\textrm{D}}\bra{\textrm{D}}$ here stands for the decoupled "dark" state, which does not evolve under the action of the drives. The transformation from the basis $(\ket{0},\ket{1},\ket{2})$ to the basis $(\ket{\textrm{D}},\ket{1},\ket{\textrm{B}})$ reads
\begin{equation}
\hat T=
\begin{pmatrix}
-b^*&0&a^*\\0&1&0\\a&0&b
\end{pmatrix},
\label{basis_change}
\end{equation}
and the time evolution unitary of the Eq. (\ref{unitary_D1B}) in the basis $(\ket{0},\ket{1},\ket{2})$ can be found as
\begin{equation}
\hat{U}=\hat{T}^{-1}\hat{\tilde{U}}\hat{T}=
\begin{pmatrix}
\abs{b}^2-\abs{a}^2&0&-2ba^*\\0&-1&0\\-2ab^*&0&-(\abs{b}^2-\abs{a}^2)
\end{pmatrix}.
\label{unitary_012}
\end{equation}

It is convenient to introduce a parameterization $a=\sin{(\theta/2)}\ex{i\phi}, b=-\cos{(\theta/2)}$ with which the time evolution can be finally written in the subspace $(\ket{0},\ket{2})$ as 
\begin{equation}
\label{gate_operator}
\hat{U}(\theta,\phi) = 
\begin{pmatrix}
\cos\theta&\ex{-i\phi}\sin\theta\\\ex{i\phi}\sin\theta&-\cos\theta
\end{pmatrix}=\vec{n}\cdot\hat{\vec{\sigma}},
\end{equation}
where the vector $\vec{n}=(\sin\theta\cos\phi,\sin\theta\sin\phi,\cos\theta)$ is the unit vector setting the direction of the rotation axis for the gate in the $(\ket{0},\ket{2})$ subspace, and $\hat{\vec{\sigma}}=(\hat{\sigma}_x,\hat{\sigma}_y,\hat{\sigma}_z)$. 

The noncommutativity of these gates can be seen from the Eq. (\ref{gate_operator}) as 
\begin{equation}
\label{noncommutativity}
[\vec{n}_1\cdot\hat{\vec{\sigma}},\vec{n}_2\cdot\hat{\vec{\sigma}}]=2i(\vec{n}_1\times\vec{n}_2) \hat{\vec{\sigma}}\ne 0
\end{equation} 
for $\vec{n}_1\ne\vec{n}_2$, where $\vec{n}_1\times\vec{n}_2$ is the vector product.

In addition to the geometric gates we use a drive $\Omega^{02}(t)$ on the $0 - 2$ transition (see Fig. \ref{three_levels}) realized via a two-photon process. This allows us to obtain information about the coherence of the state created by the geometric gate in the $(\ket{0},\ket{2})$ subspace, and also to prepare the desired initial state before acting on it with the geometric gate. Alternatively, one could use for this a sequence of pulses $0 - 1$ and $1 - 2$. Bypassing the state $\ket{1}$ with two-photon pulses has the advantage of simplifying the operation and the disadvantage is the large power levels used, which produce cross-coupling effects in weakly anharmonic systems such as the transmon.

\section{Experimental methods}

\subsection{Sample characteristics.}

\begin{figure}[h!]
\includegraphics[width=1\textwidth]{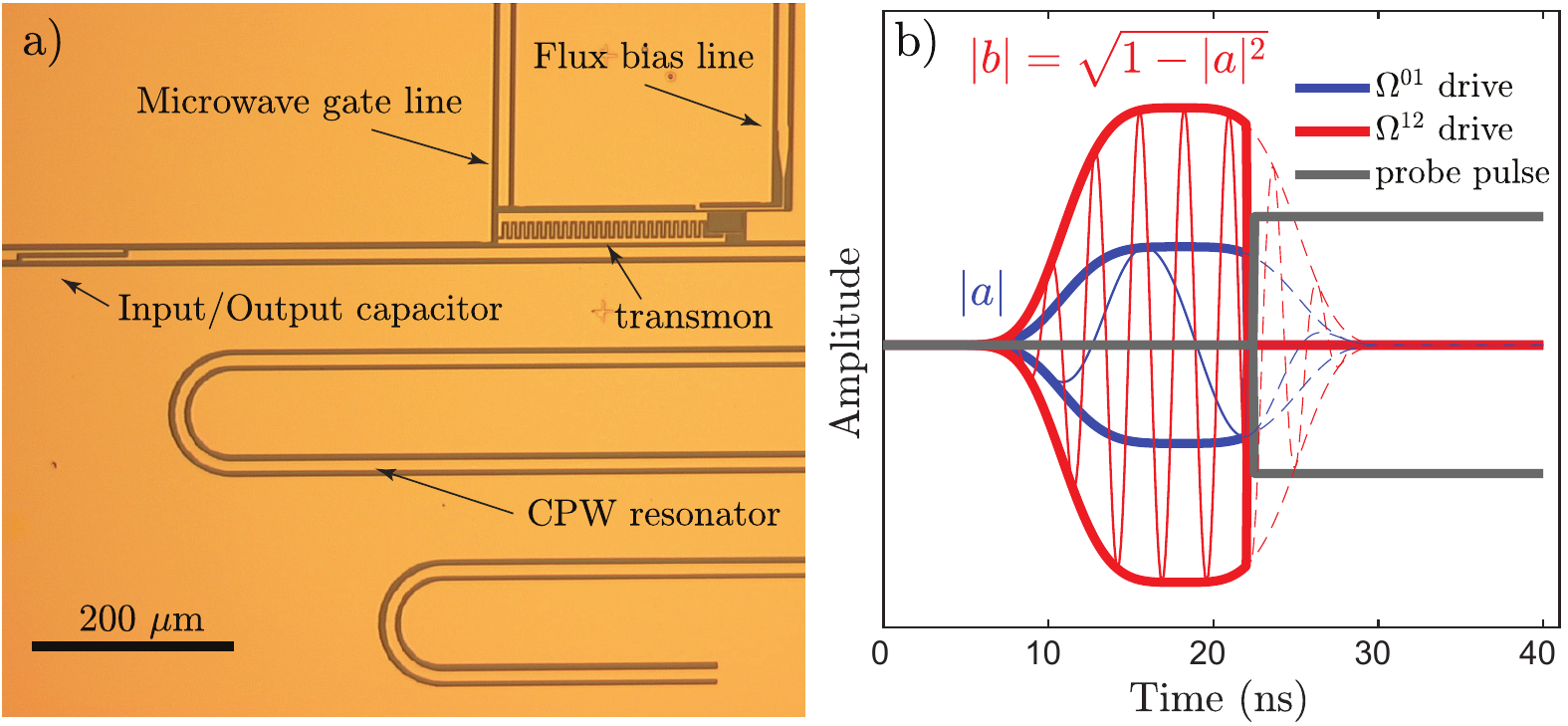}
\caption{{\bf a)} Micrograph of the sample. The aluminum film is shown in beige color, dark grey is the silicon substrate. The coplanar waveguide resonator coupled to the transmon has the input/output capacitor at one end and is grounded at the other end. The transmon has two control lines: the flux bias line and the microwave gate line. {\bf b)} Schematic of the geometric gate pulses used to manipulate the state in the $(\ket{0},\ket{2})$ subspace together with the probe pulse used to read out the state of the transmon. When the scaling parameter $\abs{a}$ of the $\Omega^{01}$ drive on the $0 - 1$ transition increases, the scaling parameter $\abs{b}$ of the $\Omega^{12}$ drive of the $1 - 2$ transition decreases as $\abs{b} = \sqrt{1 - \abs{a}^2}$.}
\label{sample_and_pulses}
\end{figure}

In our experiment we use a sample made of $100\ \textrm{nm}$ thick aluminum film deposited on top of a pure silicon substrate. The sample has a transmon~\cite{Koch} capacitively coupled to a $\lambda/4 - $ wavelength coplanar waveguide (CPW) resonator (see Fig. \ref{sample_and_pulses} a)). There are two separate onchip control lines: one of them allows to change the magnetic flux threading the transmon SQUID loop, and the other one serves for quantum state manipulation by carrying the microwave pulses to the transmon. The anharmonicity of the transmon is $E_C/h\simeq 291\ \textrm{MHz}$, determined from the difference between the frequencies of the first two transitions. The second transition is observed in the spectrum by increasing the transmon excitation power. The experiment is done at a magnetic flux value where the resonator frequency is $f_r\simeq5.1249\ \textrm{GHz}$, and the transition frequencies are $f_q^{01}\simeq7.529\ \textrm{GHz}$ and $f_q^{12}\simeq7.238\ \textrm{GHz}$. The resonator loaded quality factor is $Q\simeq7000$ and the coupling strength between the transmon and the resonator is $g/2\pi\simeq103\ \rm{MHz}$, determined from the spectroscopy measurements of the resonator around the avoided crossing region. At the "sweet spot" ($f_q^{01}\simeq7.633\ \textrm{GHz}$) the relaxation time of the qubit is $T_1 \ \simeq430\ \rm{ns}$, and the dephasing time $T_2\ \simeq 250\ \rm{ns}$ (determined from Ramsey interference). The sample is cooled down to $\sim20\ \textrm{mK}$ temperature at the mixing chamber plate of the dilution refrigerator.

\subsection{State populations measurement.}

To determine the prepared state of the three-level system we send a $2\ \mu\rm{s}$-long rectangular probe pulse to the CPW resonator input after the state preparation pulses. The signal reflected back from the resonator is downconverted in the homodyne detection scheme to slowly varying $I$ and $Q$ signals, which are recorded and used to deduce the state of the system.

We drive Rabi oscillations, sweeping the amplitudes and frequencies of the transmon excitation pulses on the $0 - 1$ and $0 - 2$ (via two-photon process) transitions and determine the $\pi$-pulses on each of these two transitions. These $\pi$-pulses are used to calibrate the system. In the calibration procedure we record the resonator responses $I_i(\tau), Q_i(\tau), i = \{0,1,2\}$ when the transmon is left unexcited, and after the application of $\pi^{01}$- and $\pi^{02}$-pulses to the transmon. Further in the experiment the resonator responses $I(t,\tau)$ and $Q(t,\tau)$ at any time $t$ can be represented as $I(t,\tau) = \sum_{i=1}^3{p_i(t)I_i(\tau)}$ (the same applies for $Q$), where $p_i$ is the population of the transmon state $\ket{i}$. This gives a way to deduce the time evolution of the three-level system populations by interrupting the state preparation pulses at some time point, sending the probe pulse to the resonator, and measuring the responses from the resonator at this time point.

\subsection{Transmon driving pulses.}

We use frequency mixing with a specific modulation frequency for each of the transitions $0 - 1$, $1 - 2$, and $0 - 2$, and a common LO frequency source to create the driving fields. As a result we have control over the relative phases of the drive pulses targeting the different transitions, their frequencies and amplitudes. The final shape of the pulses is     

\begin{equation}
\label{pulses_shapes}
\Omega^{ij}(t) = \Omega_0^{ij}\exp{\left[-\frac{1}{2}\left(\frac{t-t_0^{ij}}{t_d^{ij}}\right)^4\right]}\cos(\omega_d^{ij} t-\phi^{ij}),
\end{equation}
where $\Omega_0^{ij}$ is the Rabi amplitude of the pulse addressing the transition $i-j$, $t_0^{ij}$ - the time at which the pulse has its maximum, $t_d^{ij}$ - the time constant, determining the duration of the pulse, $\omega_d^{ij}$ - the frequency of the drive field, $\phi^{ij}$ - the phase of the drive field, and $(i,j)$ can be $(0,1)$, $(0,2)$ or $(1,2)$. The pulses are truncated at $\abs{t-t_0^{ij}} = 2t_d^{ij}$, which does not lead to the appearance of abrupt rising and falling edges, as the voltage drops much faster for this pulse shape in comparison with the Gaussian pulses. In the experiment we use $t_d^{01} = t_d^{12} = 6.5\ \rm{ns}$, and $t_d^{02} = 9\ \rm{ns}$.
In the following, for convenience, the amplitudes of the $\Omega^{ij}$ pulses will be given in volts, corresponding to the applied value from our arbitrary waveform generator.

\subsection{Pulses for the geometric gates.}

Fig. \ref{sample_and_pulses} b) shows the procedure mentioned above for the geometric gate pulses used for state preparation.

 To realize the Hamiltonian in Eq. (\ref{hamiltonian_012}) the amplitudes of the pulses are chosen such that $\Omega^{01}_{0} = \abs{a} \Omega_{2\pi}^{01}$ and $\Omega_{0}^{12} = \abs{b} \Omega_{2\pi}^{12}$, and the phase $\phi^{12}$ is taken zero, since it can be gauged away, so the forms of the drives are
 
 \begin{equation}
 \label{drive_1}
 \Omega^{01}(t) = \abs{a}\Omega_{2\pi}^{01}\exp{\left[-\half\left(\frac{t-t_0}{t_d}\right)^4\right]}\cos(\omega_d^{01}t-\phi^{01}),
 \end{equation}
 \begin{equation}
 \label{drive_2}
 \Omega^{12}(t) = \abs{b}\Omega_{2\pi}^{12}\exp{\left[-\half\left(\frac{t-t_0}{t_d}\right)^4\right]}\cos(\omega_d^{12}t).
 \end{equation}
 The maxima of these drives are at the same time $t_0$; their durations are the same and given by the parameter $t_d = 6.5\ \rm{ns}$. The couplings of the $0 - 1$ and the $1 - 2$ transitions to the external driving fields are different  (by a factor of $\sqrt{2}$ in the harmonic oscillator approximation). As a result, to create identical Rabi frequencies for these two transitions, satisfying the condition $\int\Omega(\tau)d\tau=2\pi$ mentioned above, different applied amplitudes are needed. These amplitudes are calibrated experimentally in separate Rabi oscillations measurements, where the amplitudes of the drive pulses are swept, so that they fulfill the condition
 \begin{equation}
 \label{condition}
 \int_{t_0-2t_d}^{t_0+2t_d}\Omega^{01}_{2\pi}\exp{\left[-\frac{1}{2}\left(\frac{t-t_0}{t_d}\right)^4\right]}dt = \int_{t_0-2t_d}^{t_0+2t_d}\Omega^{12}_{2\pi}\exp{\left[-\frac{1}{2}\left(\frac{t-t_0}{t_d}\right)^4\right]}dt = 2\pi.
 \end{equation}
 
 The drive $\Omega^{01}(t)$ has an additional phase $\phi^{01}$, which allows to change the relative phase of the drive on the $0-1$ transition with respect to the drive on the $1-2$ transition. The scaling parameters $\abs{a}$ and $\abs{b} = \sqrt{1-\abs{a}^2}$ as well as the phase factor $\phi^{01}$ can be changed, which causes the variation of the angles $\theta$ and $\phi$ in the parameterization Eq. (\ref{gate_operator}) and the modification of the produced geometric gate.
 
\subsection{Pulses for the two-photon process.}

\begin{figure}[h!]
	\includegraphics[width=\textwidth]{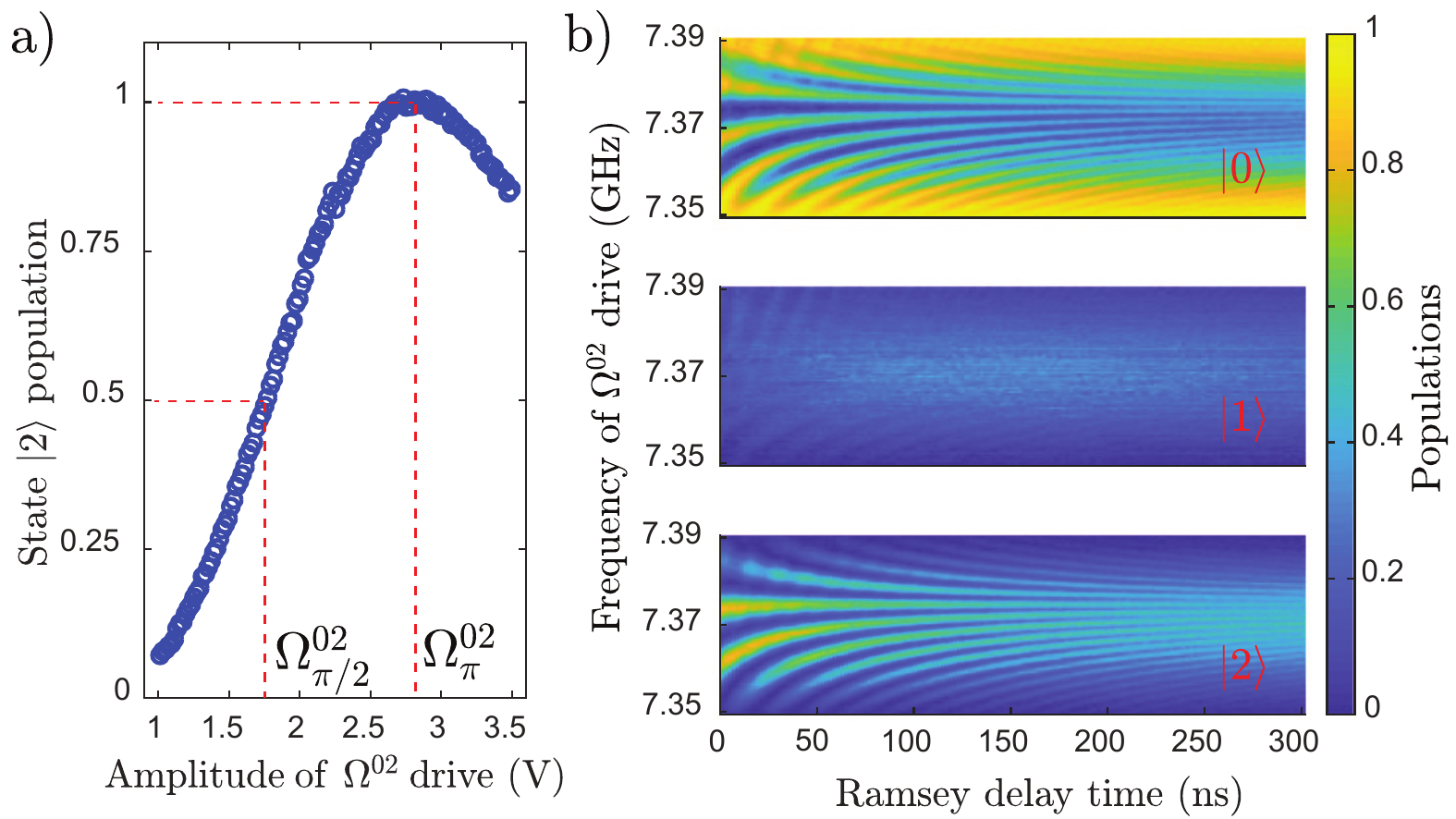}
	\caption{{\bf a)} Rabi oscillation in the $(\ket{0},\ket{2})$ subspace caused by the increase of the $\Omega^{02}_0$ amplitude of the two-photon drive. {\bf b)} Populations of the states $(\ket{0},\ket{1},\ket{2})$ in the Ramsey fringes experiment as the functions of the drive frequency and the delay time between the $\pi/2$-pulses. The state $\ket{1}$ population remains close to zero for the entire parameter space.}	
	\label{drive_02}
\end{figure}

The dipole coupling between an external driving field and the $0 - 2$ transition of the transmon is vanishingly small, so we employ a drive via a two-photon process with the increased amplitude and at the frequency around $(f_q^{01}+f_q^{12})/2$. To demostrate the state manipulation capability with two-photon pulses we measure Rabi oscillations and Ramsey fringes on the $0 - 2$ transition (Fig. \ref{drive_02}). In a Rabi oscillations experiment (Fig. \ref{drive_02} a)) we sweep the amplitude of the drive $\Omega^{02}(t)$ (Eq. (\ref{pulses_shapes})) to determine the $\pi$-pulse amplitude $\Omega^{02}_\pi$, used later for the state $\ket{2}$ calibration, and the $\pi/2$-pulse amplitude $\Omega^{02}_{\pi/2}$ with which the Ramsey fringes experiment is done. In the Ramsey fringes experiment (Fig. \ref{drive_02} b)) the time delay between the consecutive $\pi/2$-pulses and the frequency of these pulses are swept. State $\ket{1}$ population stays close to zero for the entire parameter space, which means that we indeed operate in the $(\ket{0},\ket{2})$ subspace. For $\pi/2$-rotations as used in the Ramsey experiment here and later in this work, the ideal form is
\begin{equation}
\hat{U}^{\textrm{2ph}}_{\pi/2}(\phi^{02})=\frac{1}{\sqrt{2}}
\begin{pmatrix}
1&-i\ex{2i\phi^{02}}\\
-i\ex{-2i\phi^{02}}&1
\end{pmatrix},
\label{pi2_operator}
\end{equation}
where $\phi^{02}$ is the phase of the drive Eq. (\ref{pulses_shapes}) applied to the $0 - 2$ transition. We emphasize that Eq. (\ref{pi2_operator}) gives only a simplified version of the full evolution of the system. For example, due to the fact that the two-photon pulse is strong,  ac Stark effects produce additional phase shifts which accumulate during the pulse \cite{saSTIRAPcorrection,LT28_proceedings}.

\section{Results}

 \subsection{Population control in the holonomic subspace.}
 
 We demonstrate that by varying the scaling parameters $\abs{a}$ and $\abs{b} = \sqrt{1-\abs{a}^2}$, see Eq. (\ref{drive_1},\ref{drive_2}), it is possible to manipulate the final populations of the states $\ket{0}$ and $\ket{2}$ right after the geometric gate pulses.
 
 \begin{figure}[h!]
 	\includegraphics[width=1\textwidth]{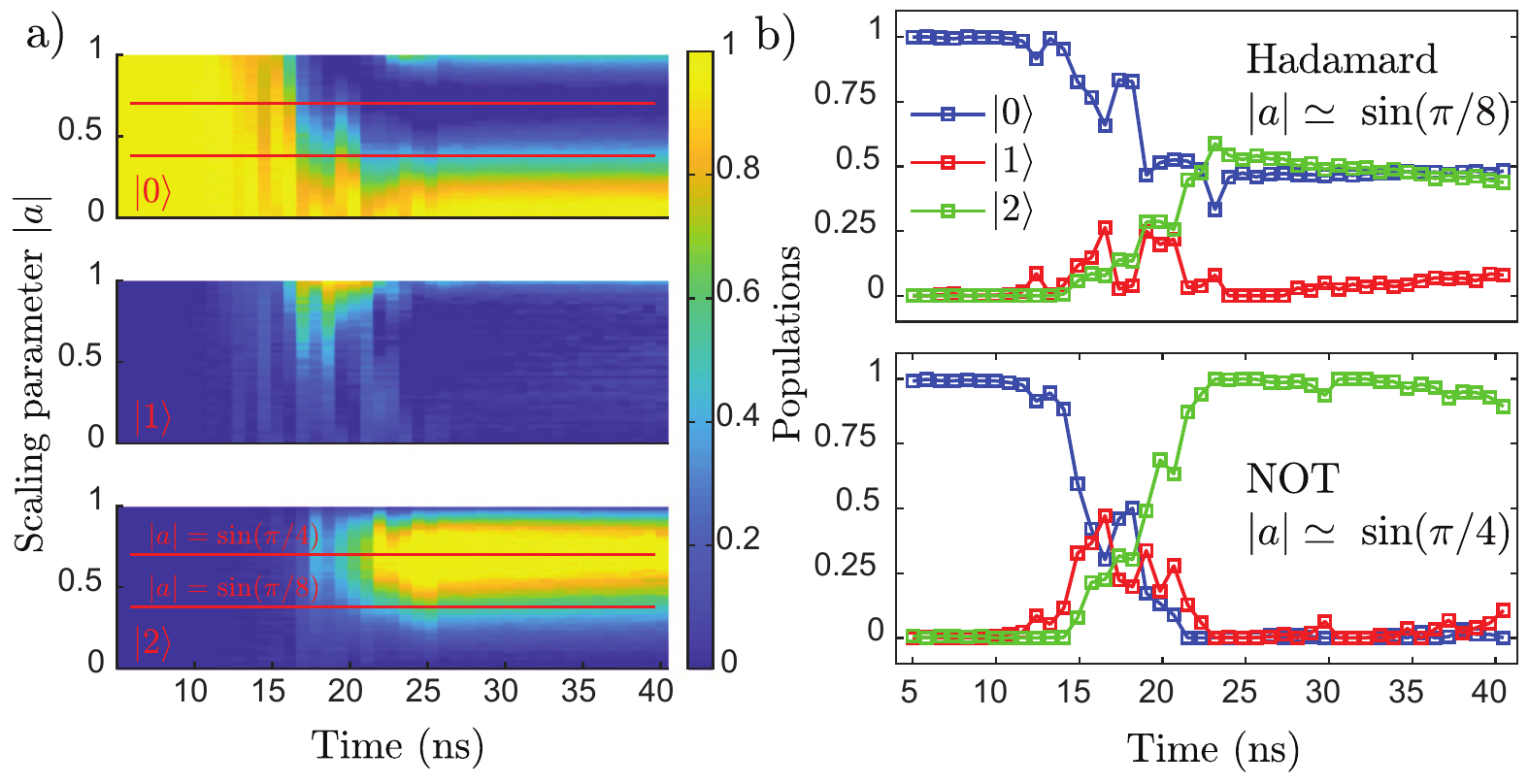}
 	\caption{{\bf a)} Time evolution of state populations during the action of the geometric gate pulses $\Omega^{01}(t)$ and $\Omega^{12}(t)$ on the transmon as a function of the scaling parameter $\abs{a}$. {\bf b)}  Evolution of the populations for the specific cases of $\theta=\pi/4$ (or $\abs{a} = \sin(\pi/8) = 0.383$) and $\theta=\pi/2$ (or $\abs{a} = \sin(\pi/4)= 0.707$) shown with the red horizontal lines in a). For the first case we create a $\pi/2$-pulse (with Hadamard gate) in the $(\ket{0},\ket{2})$ subspace, for the second case the $\pi$-pulse (with $\hat\sigma_x$ gate).}
 	\label{scaling_figure}       
 \end{figure}

Fig. \ref{scaling_figure} a) shows the populations of the three lowest transmon states during the geometric gate at different values of the scaling parameter $\abs{a}$ (different values of $\theta$ in Eq. (\ref{gate_operator})), and phase factor $\phi^{01}=\pi$. If the scaling parameter $\abs{a} = 0$, there is no drive applied to the $0 - 1$ transition, but the drive $\Omega^{12}$ has its maximal amplitude. Initially the first excited state is not populated at all, and as a result the populations stay unchanged. With the increase of the parameter $\abs{a}$ the population transfer in the $(\ket{0},\ket{2})$ subspace starts to take place, at the same time the final population of the state $\ket{1}$ is close to zero. It is possible to determine a special value of $\abs{a}$ at which we can create an equal superposition between the $\ket{0}$ and $\ket{2}$ ($\pi/2$-pulse via Hadamard gate) or implement a full population transfer to the second excited state ($\pi$-pulse via $\hat{\sigma}_x$ gate). These values are depicted with the red horizontal lines in the Fig.\ref{scaling_figure} a).  Fig. \ref{scaling_figure} b) demonstrates detailed changes of the populations at these special values of $\abs{a}=\sin(\pi/8)$ (the first plot) and $\abs{a}=\sin(\pi/4)$ (the second plot). One can see that right after the end of the drive pulses at $31\ \rm{ns}$ in the first case $p_0$ and $p_2$ are close to $1/2$, and for the second case $p_2 \simeq 1$ and the other two states are unpopulated.

We measure only populations of the states and test that the final populations reached after the end of the geometric gates do not depend on the relative phase $\phi^{01}$ (Eq. (\ref{drive_1})) between the drive pulses forming the gates.

\subsection{Phase control in the holonomic subspace.}

\begin{figure}[h!]
	\centering
	\includegraphics[width = 1.0\textwidth]{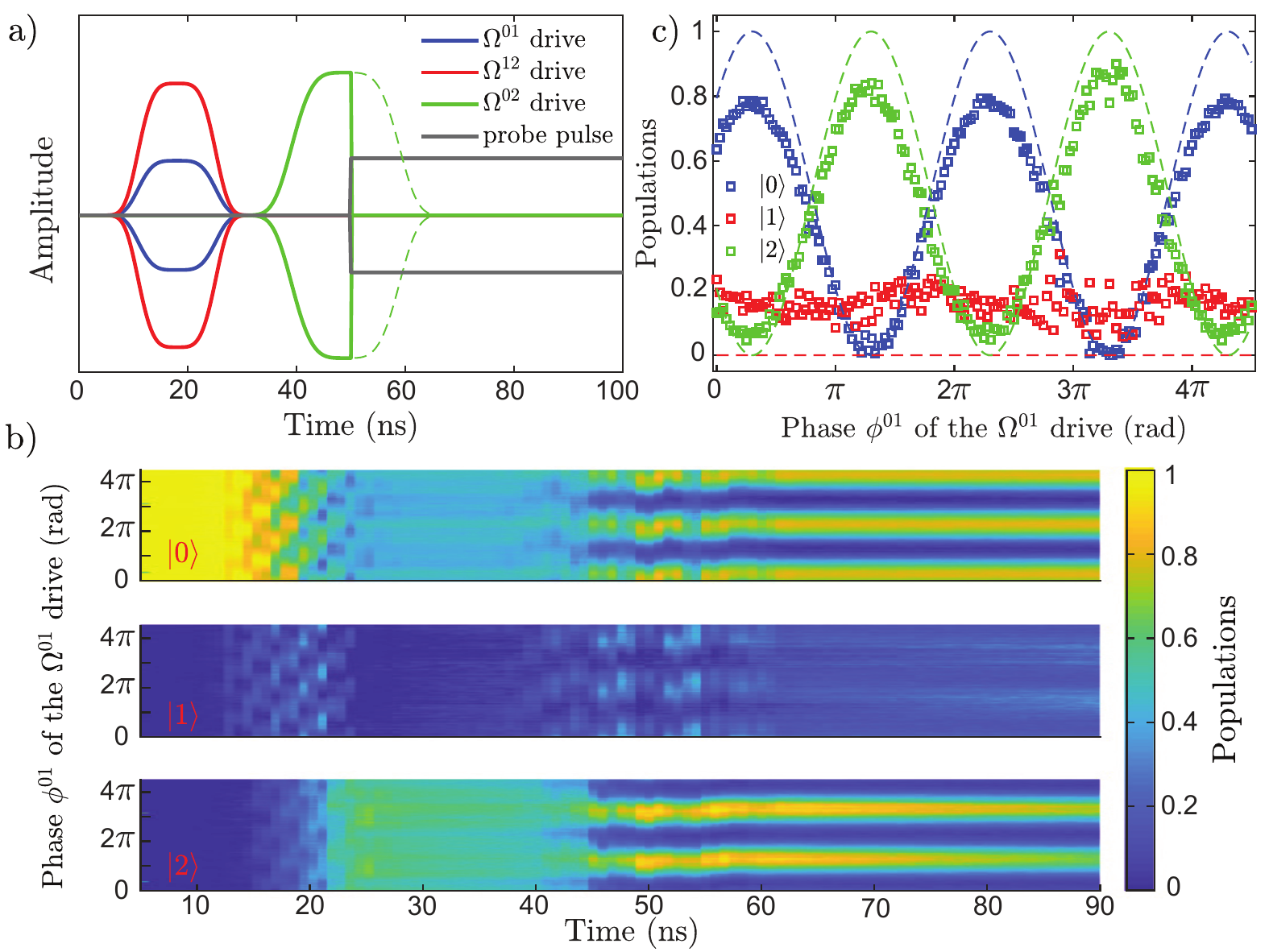}
	\caption{{\bf a)} Schematics of the pulses sent to the transmon microwave gate line and to the resonator. {\bf b)} Evolution in time of the states populations as a function of the phase factor $\phi^{01}$ of the $\Omega^{01}$ drive. {\bf c)} Phase dependence of the states populations at a time $67\ \textrm{ns}$ right after the end of the entire pulse sequence. Dashed lines represent the ideal case Eq. (\ref{HU_populations}) with a constant phase shift: blue line for the ground state, and green for the second excited state.
	}
	\label{gate_hadamard_figure}
\end{figure}

If the scaling parameters are chosen so that $\abs{a} = \sin(\pi/8)$, and $\abs{b} = \cos(\pi/8)$ (angle $\theta=\pi/4$ in the parameterization of Eq. (\ref{gate_operator})), the two drives $\Omega^{01}(t)$ and $\Omega^{12}(t)$ create the Hadamard gate. This gate applied to the ground state $\ket{0}$ forms an equal superposition $\ket{\psi}=(\ket{0}+\ex{i\phi}\ket{2})/\sqrt{2}$ between the states $\ket{0}$ and $\ket{2}$. The phase $\phi$ here can be controlled via changes in the phase factor $\phi^{01}$ of the drive pulse $\Omega^{01}(t)$, see Eq.(\ref{drive_1}). Experimentally we can only measure the populations of the states, and to reveal the phase $\phi$ of the created state, we apply a two-photon $\pi/2$-pulse on the $0 - 2$ transition right after the geometric gate pulses. The sequence of pulses is shown in Fig. \ref{gate_hadamard_figure} a). We interrupt the pulse sequence at consecutive time points and determine the population of each state. The result is shown in Fig. \ref{gate_hadamard_figure} b), where the time evolution of the populations during the entire pulse sequence is shown as a function of the phase factor $\phi^{01}$.

From  Fig. \ref{gate_hadamard_figure} b) one can see that at the time around $31\ \rm{ns}$ from the beginning of the sequence, right after the geometric gate (Fig. \ref{gate_hadamard_figure} a)), populations of the states $\ket{0}$ and $\ket{2}$ are equal and close to $1/2$ regardless of the phase factor $\phi^{01}$. At the same time the population of the state $\ket{1}$ is close to zero. This proves that the geometric gate, formed as described above and applied to the state $\ket{0}$, indeed creates an equal superposition state in the $\{\ket{0},\ket{2}\}$ subspace and can be represented in this holonomic subspace in the form of Eq. (\ref{gate_operator}) with $\theta=\pi/4$:
\begin{equation}
\hat U(\pi/4,\phi^{01})=\frac{1}{\sqrt{2}}
\begin{pmatrix}
1&\ex{-i\phi^{01}}\\\ex{i\phi^{01}}&-1
\end{pmatrix} .
\label{Hadamard}
\end{equation} 

After the end of the two-photon $\pi/2$-pulse at the time $67\ \rm{ns}$ there are oscillations in the populations (Fig. \ref{gate_hadamard_figure} c)) with $2\pi$ periodicity, which means that the phase $\phi$ of the state $(\ket{0}+\ex{i\phi}\ket{2})/\sqrt{2}$, created by the geometric gate pulses, is different for different phase factors $\phi^{01}$ of the drive, and can be controlled by tuning the latter.

The resulting state after this pulse sequence in the basis $\{\ket{0},\ket{2}\}$ can be represented as 
\begin{equation}
\label{HU_sequence}
\hat{U}^{\textrm{2ph}}_{\pi/2}(\phi^{02}=0)\hat{U}(\pi/4,\phi^{01})\ket{0}
=
\frac{1}{2}\begin{pmatrix}
1-i\ex{i\phi^{01}}\\-i+\ex{i\phi^{01}}
\end{pmatrix}.
\end{equation}

 The populations of the states $\ket{0}$ and $\ket{2}$ at the end of the sequence then read 

\begin{equation}
\label{HU_populations}
p_0(\phi^{01}) = \frac{1+\sin(\phi^{01})}{2},\quad p_2(\phi^{01}) = \frac{1-\sin(\phi^{01})}{2}.
\end{equation}

 These equations are plotted with the dashed lines in the Fig. \ref{gate_hadamard_figure} c), including a constant phase shift corresponding to nonidealities in the two-photon gate \cite{saSTIRAPcorrection,LT28_proceedings}. We attribute the reduction in the amplitude of these oscillations and the rise from zero of the state $\ket{1}$ population to the relaxation process of our transmon, which is noticeable at this time point.
 
 \subsection{Holonomic operations on arbitrary initial states}
 
 \begin{figure}[h!]
 	\centering
 	\includegraphics[width=1.0\textwidth]{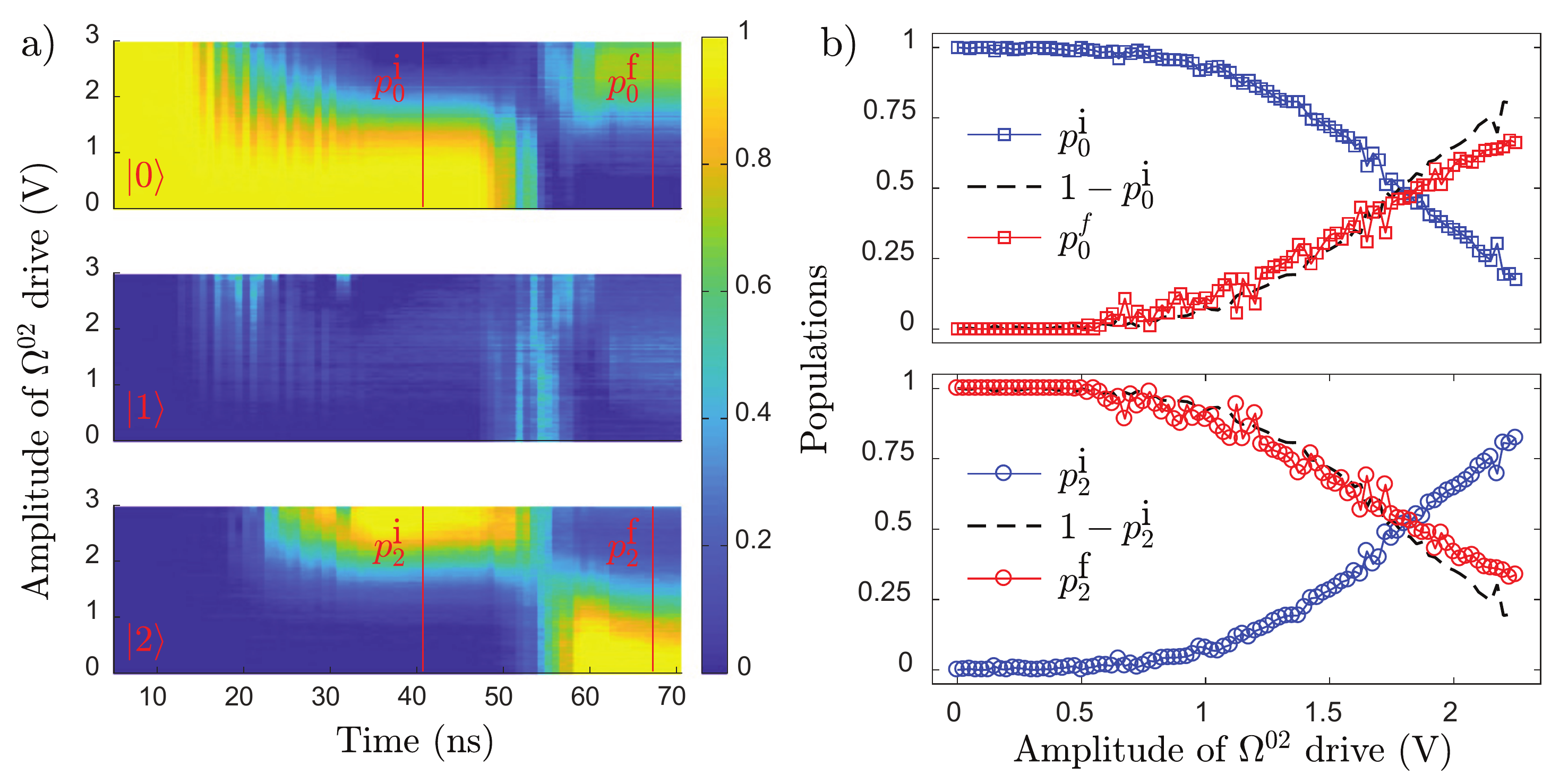}
 	\caption{{\bf a)} Time evolution of the populations of the three lowest energy states of the transmon during the action of the $\Omega^{02}$ two-photon and NOT geometric gate pulse sequence at different amplitudes of the two-photon $\Omega^{02}$ drive. Red vertical lines mark the time points at the end of the $\Omega^{02}$ drive pulse, where the initial state is prepared with populations $(p_0^\textrm{i},p_2^\textrm{i})$, and at the end of the geometric gate pulses when the populations are $(p_0^\textrm{f},p_2^\textrm{f})$. {\bf b)} Demonstration of population inversion under the action of the NOT holonomic gate applied to the different initial states in the $(\ket{0},\ket{2})$ subspace of the transmon.}
 	\label{pi_flip_figure}
 \end{figure} 

In the previous experiments we have been applying the geometric gates to the transmon in the ground state. It is of course possible to apply the holonomic rotations to any superposition of the ground state and the second excited state. To demonstrate this, we first prepare the system in a state $\alpha\ket{0}+\ex{i\phi}\beta\ket{2}$, applying a two-photon drive pulse $\Omega^{02}$ with different amplitudes and phases $\phi^{02}$. The amplitude of this pulse determines the measured populations $p_0=\abs{\alpha}^2$ and $p_2=\abs{\beta}^2$ at the end of the pulse. For the holonomic operation, when the scaling parameters are $\abs{a}=\sin(\pi/4)$ and $\abs{b}= \cos(\pi/4)$   ($\theta=\pi/2$) the resulting operation is a NOT gate in the $(\ket{0},\ket{2})$ subspace of the three-level system. The schematics of the pulse sequence is the same as in Fig. \ref{gate_hadamard_figure} a), but this time the $\Omega^{02}$ two-photon pulse goes before the simultaneous drives $\Omega^{01}$ and $\Omega^{12}$ forming the geometric NOT gate. At the end of the sequence we measure the resulting populations. Fig. \ref{pi_flip_figure} a) shows the populations of the three lowest states as a function of the time in the sequence of the state preparation pulses and the amplitude of the drive $\Omega^{02}$. Time slices at the end of the two-photon $\Omega^{02}$ pulse ($41\ \textrm{ns}$), and at the end of the geometric gate ($67\ \textrm{ns}$) are shown with red vertical lines in the figure.  
 
 One can see that with the rise of the $\Omega^{02}$ amplitude the populations of the states $\ket{0}$ and $\ket{2}$ at the end of the two-photon pulse interchange (see blue data points in the Fig. \ref{pi_flip_figure} b) for $p_0^\textrm{i}$ and $p_2^\textrm{i}$). The black dashed lines in the Fig. \ref{pi_flip_figure} b) represent the expectation values for $p_0$ and $p_2$ after the action of the NOT gate: $1-p_0^\textrm{i}$ and $1-p_2^\textrm{i}$. The red data points show the populations $p_0^\textrm{f}$ and $p_2^\textrm{f}$ right after the end of the geometric gate pulses, and they coincide well with the expected values. To take into account the effects of relaxation in the Fig. \ref{pi_flip_figure} b) we have corrected the state $\ket{2}$ population by the measured values of state $\ket{1}$ population at the corresponding times, extracted from Fig. \ref{pi_flip_figure} a). At higher amplitudes of the two-photon drive $\Omega^{02}$ the higher energy levels of the transmon start to be excited, and the inversion of populations between $\ket{0}$ and $\ket{2}$ does not occur with high fidelity.
 
 
 \begin{wrapfigure}[21]{l}{0.48\textwidth}
 	\centering
 	\includegraphics[width=0.48\textwidth]{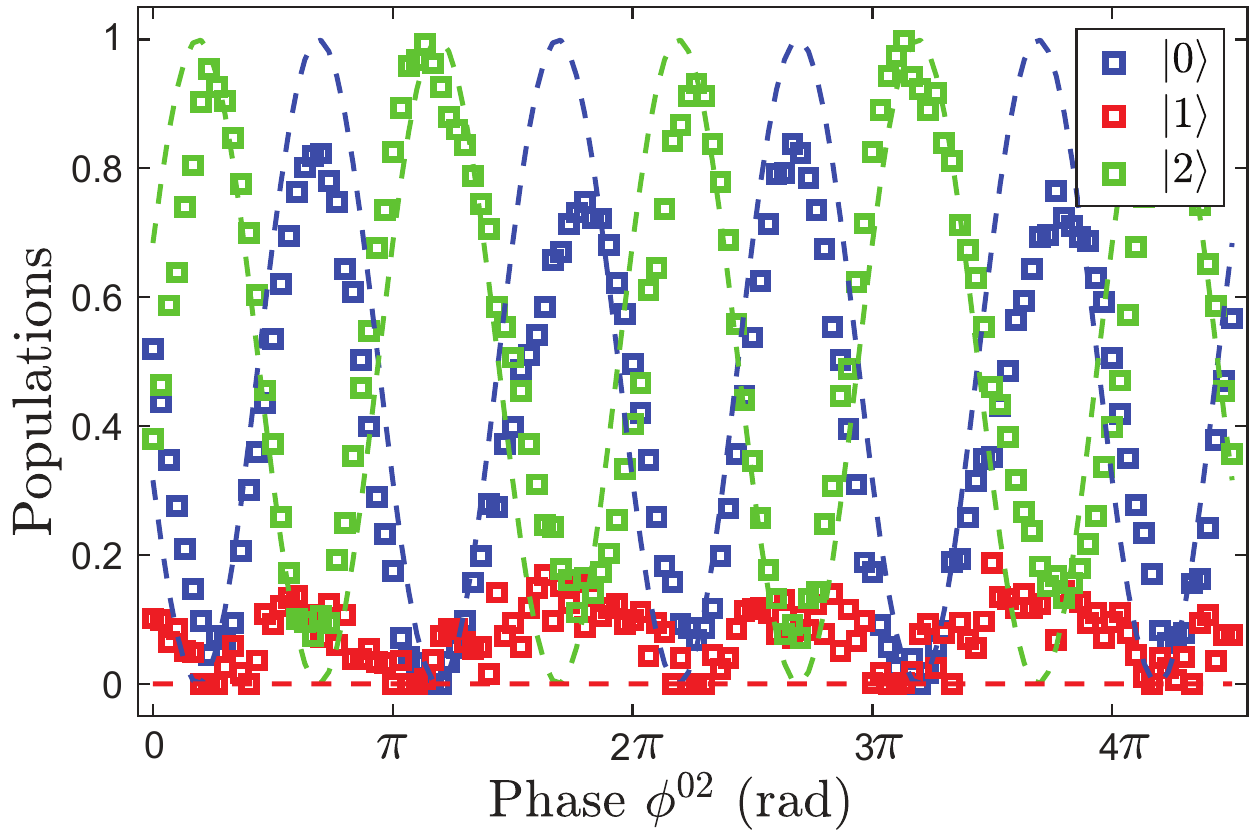}
 	\caption{Dependence of the three lowest transmon states populations on the phase $\phi^{02}$ of the two-photon $\pi/2$-pulse right after the end of the Hadamard gate, created with the geometric gate pulses. Dashed lines show the expected behaviour, given by the Eq. (\ref{pi2_hadamard_populations}), blue line for the ground state and green line for the second excited state.}
 	\label{pi2_hadamard_figure}
 \end{wrapfigure}
 
 We also test the action of the Hadamard gate created with the geometric gates on the equal superposition states $(\ket{0}+\ex{i\phi}\ket{2})/\sqrt{2}$ with different phases $\phi$. This time the geometric gate again follows the $\pi/2$-pulse, but the phase $\phi^{02}$ of the $\pi/2$-pulse is swept, and the geometric gate parameters are $\abs{a}=\sin{(\pi/8)}, \abs{b}=\cos{(\pi/8)}$, $\phi^{01}=0$ (corresponding to $\theta=\pi/4$), which produces a Hadamard gate. The populations of the three lowest states of the transmon right after the end of the Hadamard gate are shown in Fig. \ref{pi2_hadamard_figure} for different values of $\pi/2$-pulse phase $\phi^{02}$. Note that the populations oscillate with $\pi$ periodicity, which is the manifestation of the two-photon process used to create the initial state $(\ket{0}+\ex{i\phi}\ket{2})/\sqrt{2}$.  
 
 The result can be understood as a consecutive action on the ground state $\ket{0}$ first with the $\hat{U}^{\textrm{2ph}}_{\pi/2}(\phi^{02})$ gate having the phase $2\phi^{02}$, and then with the Hadamard gate $\hat{U}(\pi/4,0)$.
 
 \begin{equation}
 \hat{U}(\pi/4,0)\hat{U}^{\textrm{2ph}}_{\pi/2}(\phi^{02})\ket{0}=\frac{1}{2}
 \begin{pmatrix}
 1-i\ex{-2i\phi^{02}}\\ 1+i\ex{-2i\phi^{02}}
 \end{pmatrix}.
 \label{pi2_hadamard_equation}
 \end{equation}
 
 The resulting populations are
 \begin{equation}
 p_0(\phi^{02})=\frac{1-\sin{(2\phi^{02})}}{2},\quad p_2(\phi^{02})=\frac{1+\sin{(2\phi^{02})}}{2}.
 \label{pi2_hadamard_populations}
 \end{equation}
 
 The populations of the Eq. (\ref{pi2_hadamard_populations}) are plotted in Fig. \ref{pi2_hadamard_figure} (up to a constant phase) with the blue dashed line for the ground state and green dashed line for the second excited state.

 \section{Conclusions}
  
 We experimentally demonstrate on a fully coplanar superconducting structure with the transmon artificial atom that the geometric gates can be used to manipulate the quantum state of the three-level system. The geometric gates are created by applying simultaneously two microwave drives to the $0 - 1$ and $1 - 2$ transitions of the transmon. We show that it is possible to control the populations of the final state in the $(\ket{0},\ket{2})$ subspace as well as the phase of this state. The populations are given by the amplitudes ratio between the drive pulses, the phase - by the relative phase factor of one of the drives with respect to another one. We also test the operation of the NOT gate and Hadamard gate, created with the geometric gate pulses, acting with them on different initial superposition states in the $(\ket{0},\ket{2})$ subspace. To visualize the phase of the state produced by the geometric gate and to prepare the initial states for the NOT gate and Hadamard gate characterization we employ a two-photon drive pulse on the $0 - 2$ transition. 
 
%
%

\ 

\newpage
{\bf Acknowledgments}

We are very grateful for the financial support from the Academy of Finland (project 263457 and Center of Excellence "Low Temperature Quantum Phenomena and Devices" - project 250280), the Center for Quantum Engineering at Aalto University (project QMET and QMETRO), and V\"{a}isal\"{a} Foundation. The experiments were performed at the cryogenic facilities of the Low Temperature Laboratory at Aalto University.




\end{document}